\documentclass[apj,twocolumn]{openjournal}
\usepackage{newtxtext,newtxmath,enumitem,multirow,ctable}
\usepackage[T1]{fontenc}
\usepackage[breaklinks,colorlinks,citecolor=blue,urlcolor=blue]{hyperref}

\newcommand{\Alf}{{Alfv\'en}}

\newcommand{\paperone}{Paper {\small I}}
\newcommand{\papertwo}{Paper {\small II}}

\newcommand{\orcidauthor}[3]{\author{\href{http://orcid.org/#1}{#2$^{#3}$}}}

\shorttitle{Flux-Frozen, Hyper-Magnetized Disks}
\shortauthors{Hopkins et al.}

\begin{document}

\title{\vspace{-0.8cm}An Analytic Model For Magnetically-Dominated Accretion Disks\vspace{-1.5cm}}

\orcidauthor{0000-0003-3729-1684}{Philip F. Hopkins}{1,*}
\orcidauthor{0000-0001-8479-962X}{Jonathan Squire}{2}
\orcidauthor{0000-0001-9185-5044}{Eliot Quataert}{3}
\orcidauthor{0000-0002-8659-3729}{Norman Murray}{4}
\orcidauthor{0000-0003-1598-0083}{Kung-Yi Su}{5}
\orcidauthor{0000-0001-8867-5026}{Ulrich P. Steinwandel}{6}
\orcidauthor{0000-0002-4086-3180}{Kyle Kremer}{1}
\orcidauthor{0000-0002-4900-6628}{Claude-Andr\'{e} Faucher-Gigu\`{e}re}{7}
\orcidauthor{0000-0002-3977-2724}{Sarah Wellons}{8}

\affiliation{$^{1}$TAPIR, Mailcode 350-17, California Institute of Technology, Pasadena, CA 91125, USA}
\affiliation{$^{2}$Physics Department, University of Otago, 730 Cumberland St., Dunedin 9016, New Zealand}
\affiliation{$^{3}$Department of Astrophysical Sciences, Princeton University, Princeton, NJ 08544, USA}
\affiliation{$^{4}$Canadian Institute for Theoretical Astrophysics, University of Toronto, Toronto, ON M5S 3H8, Canada}
\affiliation{$^{5}$Black Hole Initiative, Harvard University, 20 Garden St, Cambridge, MA 02138, USA}
\affiliation{$^{6}$Center for Computational Astrophysics, Flatiron Institute, 162 5th Ave., New York, NY 10010 USA}
\affiliation{$^{7}$CIERA and Department of Physics and Astronomy, Northwestern University, 1800 Sherman Ave, Evanston, IL 60201, USA}
\affiliation{$^{8}$Department of Astronomy, Van Vleck Observatory, Wesleyan University, 96 Foss Hill Drive, Middletown, CT 06459, USA}

\thanks{$^*$E-mail: \href{mailto:phopkins@caltech.edu}{phopkins@caltech.edu}},

\begin{abstract}
Recent numerical cosmological radiation-magnetohydrodynamic-thermochemical-star formation simulations have resolved the formation of quasar accretion disks with Eddington or super-Eddington accretion rates onto supermassive black holes (SMBHs) down to a few hundred gravitational radii. These ``flux-frozen'' and hyper-magnetized disks appear to be qualitatively distinct from classical $\alpha$ disks and magnetically-arrested disks: the midplane pressure is dominated by toroidal magnetic fields with plasma $\beta \ll 1$ powered by advection of magnetic flux from the interstellar medium (ISM), and they are super-sonically and trans-\Alf{ically} turbulent with cooling times short compared to dynamical times yet remain gravitationally stable owing to magnetic support. In this paper, we present a simple analytic similarity model for such disks. For reasonable assumptions, the model is entirely specified by the boundary conditions (inflow rate at the BH radius of influence [BHROI]). We show that the scalings from this model are robust to various detailed assumptions, agree remarkably well with the simulations (given their simplicity), and demonstrate the self-consistency and gravitational stability of such disks even in the outer accretion disk (approaching the BHROI) at hyper-Eddington accretion rates.
\end{abstract}


\maketitle

\section{Introduction}
\label{sec:intro}

Accretion disks play a crucial role in many fields of astrophysics. In particular around supermassive BHs, these disks power active galactic nuclei (AGN) and quasars, and are believed to supply mass at rates exceeding $\gtrsim 10\,{\rm M_{\odot}\,yr^{-1}}$ to the BH, ultimately building up most of the mass in SMBHs \citep{schmidt:1963.qso.redshift,soltan82}. The vast majority of the theoretical literature on quasar accretion disks has assumed as a starting point something like the \citet[][hereafter SS73]{shakurasunyaev73} $\alpha$-disk model, which assumes that disks are geometrically thin (height $H \ll R$), optically thick (black-body like), sub-sonically turbulent (sonic Mach number $\mathcal{M}_{s} < 1$), slowly cooling ($t_{\rm cool} \gg t_{\rm dyn} = 1/\Omega$), gas and/or radiation-pressure-dominated (plasma\footnote{Throughout, we adopt Gaussian units, and use the convention $\beta \equiv c_{s}^{2}/v_{A}^{2}$ for convenience, but note this differs by a factor of $\sim 2$ (depending on the gas equation-of-state) from the common $\beta \equiv P_{\rm gas}/P_{\rm mag}$.} $\beta \equiv c_{s}^{2}/v_{A}^{2} > 1$), radiatively efficient, well-ionized, and can be parameterized by an effectively constant-$\alpha$ viscosity where $\alpha \sim (\delta v)^{2}/c_{s}^{2} < 1$ represents some Maxwell or Reynolds stresses and the kinematic viscosity scales as $\nu_{\rm visc} \equiv \alpha c_{s} ^{2}/\Omega$. While many variations have been introduced (including radiatively inefficient, advection-dominated, magnetically-arrested, magnetically-elevated, ``slim,'' and gravito-turbulent disks; for reviews see \citealt{frank:2002.accretion.book,abramowicz:accretion.theory.review}), for luminous quasars some form of SS73-like analytic model is still most often the ``default'' reference. 

Recently, \citet[][\paperone]{hopkins:superzoom.overview} and \citet[][\papertwo]{hopkins:superzoom.disk} presented the first simulations to self-consistently follow gas in a cosmological simulation from $>$\,Mpc to $<100\,$au scales (a few hundred gravitational radii) around an accreting SMBH, including the physics of magnetic fields (seeded from trace cosmological values), multi-band radiation-hydrodynamics, non-equilibrium multi-phase radiative thermo-chemistry and cooling, self-gravity, star formation, and stellar evolution/feedback (jets, stellar mass-loss, radiation, supernovae). In these simulations, gas around the BHROI\footnote{Defined as the radius interior to which the BH dominates the potential over its host galaxy of characteristic velocity dispersion $\sigma_{\rm gal}$, or $R_{\rm BHROI} \sim G\,M_{\rm bh}/\sigma_{\rm gal}^{2} \sim 5\,{\rm pc}$ in the reference simulations.} is tidally captured by the SMBH of mass $M_{\rm bh} \sim 10^{7}\,M_{\odot}$ from larger-scale ISM gas complexes in the galaxy, and free-falls briefly before circularizing to form an accretion disk with Toomre $Q \sim v_{\rm support}\,\Omega/\pi\,G\,\Sigma_{\rm gas} \gg 1$ and little to no star formation or fragmentation on sub-pc scales. This disk evolves in quasi-steady-state and sustains super-Eddington accretion (up to $\dot{M} \sim 20-30\,{\rm M_{\odot}\,yr^{-1}}$) onto the SMBH for at least $\sim 10^{5}$ dynamical times at the inner simulation boundary (the simulation duration). Crucially, in \papertwo\ where the disk properties were studied in detail, it was shown that the disks are strongly magnetized, with $\beta \sim 10^{-4}-10^{-2}$ in the midplane, in the form of primarily toroidal magnetic field (but with mean radial fields and quasi-isotropic turbulent fields only a factor of a few less strong) owing to amplification of magnetic flux accreted from the ISM. These stabilize the disk against catastrophic fragmentation and star formation: without magnetic fields, the disks were shown to be orders-of-magnitude less massive and support factor of $\sim 1000$ lower accretion rates and higher star formation rates. The disks also have a flared structure ($H/R \sim 0.1-0.3$ at large radii) with weak vertical stratification owing to trans-\Alf{ic},  highly super-sonic turbulence, which is sustained by rapid cooling ($\mathcal{M}_{A} \sim \delta v / v_{A} \sim 1$, with $\mathcal{M}_{s}^{2} \sim 1/\beta \sim 1/t_{\rm cool}\,\Omega \gg 1$). 

These disks are therefore qualitatively distinct from SS73-like $\alpha$-disks in most respects. As discussed in \papertwo, they also appear to qualitatively differ in key respects from most or all historical models of ``strongly magnetized'' disks such as magnetically ``elevated'' or ``arrested'' disks \citep[e.g.][]{bisnovatyi.kogan:1976.mad.disk,miller.stone:2000.magnetically.elevated.disk,narayan:2003.mad.disk}, hence referring to them as ``flux-frozen'' and ``hyper-magnetized.'' And their properties may resolve some crucial long-standing questions: most obviously, the well-known problem that an SS73-like disk with quasar luminosities should be violently gravitationally unstable outside of just a couple hundred gravitational radii \citep{goodman:qso.disk.selfgrav}. 

But, while simulations like those in \papertwo\ are ultimately necessary to explore how, when, and ``what kind'' of accretion disks form (let alone to follow the highly non-linear, multi-physics evolution), they are complex and difficult to use for simple physical intuition and interpretation. Moreover, owing to computational expense, the simulations capture just one system at one moment in cosmic time, so it is not obvious what we can reliably extrapolate to other systems with different BH masses or inflow rates. Motivated by these considerations, in this manuscript (\S~\ref{sec:analytic}) we develop a simple self-similar analytic model for the super-sonically turbulent, rapidly-cooling, hyper-magnetized/flux-frozen disks seen in the simulations (building on previous work such as \citealt{pariev:2003.mag.dominated.disk.models}), modeling their density, magnetic, and angular momentum structure (\S~\ref{sec:analytic.structure}), as well as thermal and radiation structure (\S~\ref{sec:analytic.thermal}), and validating this against the direct simulations. We discuss the sensitivity (or lack thereof) of the results to different assumptions (\S~\ref{sec:analytic.caveats}) and reasonable expectations for the model parameters, then compare to the scalings predicted by the classic SS73 disk theory (\S~\ref{sec:analytic.ss73}), as well as some previous, more similar literature models for ``magnetically dominated'' disks (\S~\ref{sec:analytic.bp07}) and magnetically elevated or arrested disks (\S~\ref{sec:analytic.arrested}). We discuss the upper limits to accretion and likely limitations of the model at small radii (approaching the horizon) in \S~\ref{sec:analytic.eddington}, and summarize in \S~\ref{sec:conclusions}. Note that our primary focus here is on quasar accretion disks around SMBHs, as opposed to stellar-mass BH binaries (where it is less clear our assumptions apply).

\begin{figure*}
	\centering
	\includegraphics[width=0.99\textwidth]{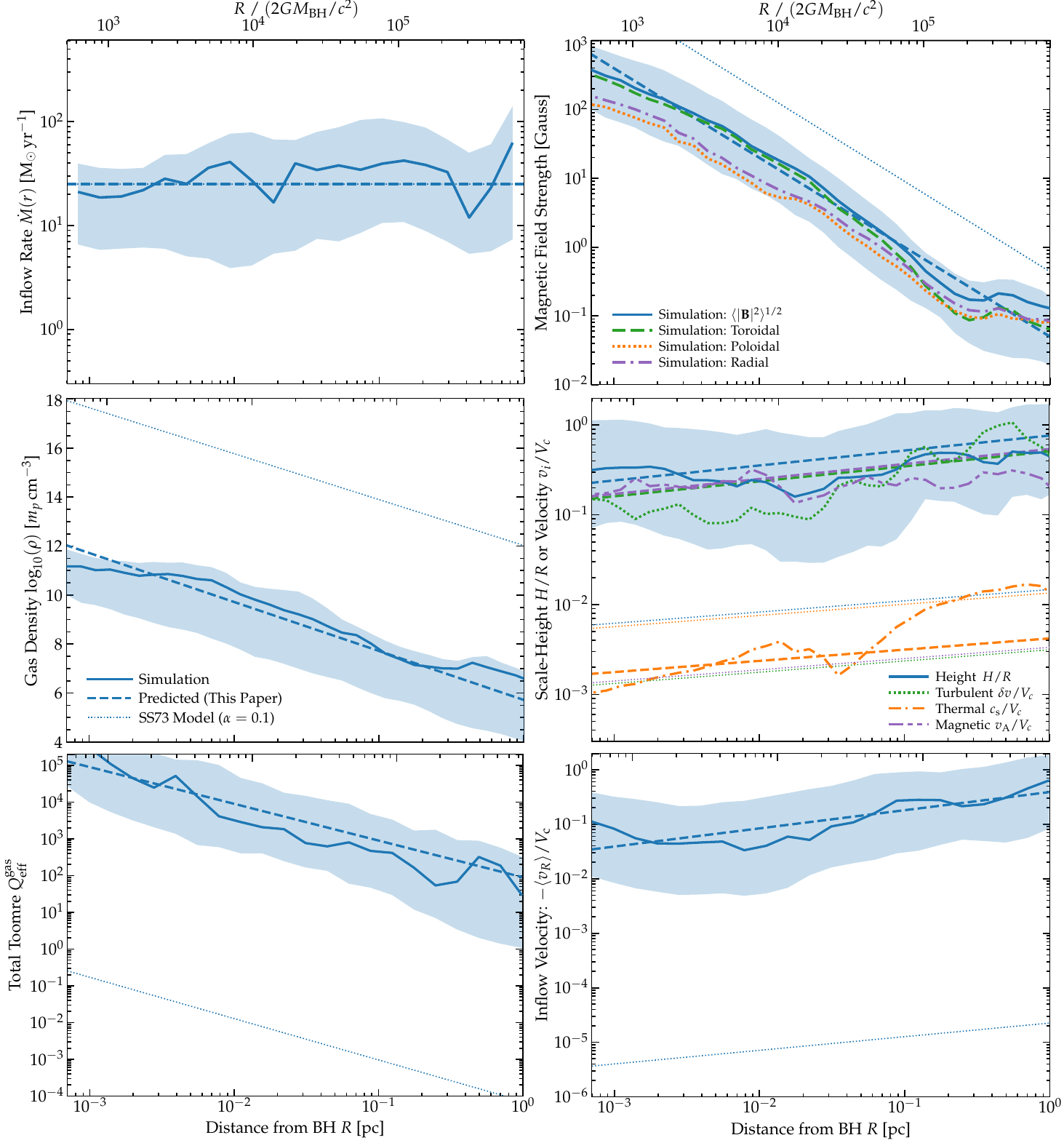}
	\caption{Comparison of the model predictions here ({\em thick dashed} lines, \S~\ref{sec:analytic}) to the numerical simulations of quasar accretion disks forming from cosmological initial conditions in \papertwo\ ({\em solid} lines show mean values at each $R$ at one moment/snapshot, shaded range shows the $90\%$ inclusion interval for all gas at that $R$, averaged over all simulation snapshots). We compare inflow rate $\dot{M}(r)$ (here {\em net} inflow less outflow); magnetic field strength (with the simulation example showing the toroidal/poloidal/radial decomposition); gas density; scale height $H/R$; turbulent (1D direction-averaged rms $\delta v$), thermal sound ($c_{s}$) and \Alf\ ($v_{A}$) speeds; mean inflow velocity $-\langle v_{R}\rangle$; and effective Toomre $Q$ parameter (including thermal, magnetic, and turbulent support). Simulation quantities are volume-averaged in the midplane ($|z|<H$), and plotted versus cylindrical radius from the SMBH $R$, from $\sim 80\,$au (the inner simulation accretion boundary) to $\sim 1\,$pc (exterior to which the physics become more ISM-like, with significant star formation in a GMC-like complex which is being tidally disrupted by the SMBH to form the accretion disk, biasing the statistics; see \paperone). For the models, we take the parameters $M_{\rm bh}=1.3\times10^{7}\,M_{\odot}$, $\dot{M}_{\rm bh} \approx 25\,{\rm M_{\odot}\,yr^{-1}}$ (plotted; $\dot{m}\approx 100$), and $r_{\rm ff} \approx R_{\rm BHROI} \approx 5\,{\rm pc}$ directly from the simulation, and assume trans-\Alf{ic} turbulence ($\psi_{A}=1$, $\zeta=1/3$). We contrast a \citet[][SS73]{shakurasunyaev73} model with $\alpha=0.1$ ({\em thin dotted}). The analytic model reproduces the key simulation properties reasonably well, especially compared to an SS73 model which predicts quantities like $Q$ differing by factors as large as $\sim 10^{8}$.
	\label{fig:sim.model.comparison}}
\end{figure*}

\section{The Model}
\label{sec:analytic}

With this motivation in mind, we develop here a simple approximate analytic similarity model for the ``hyper-magnetized'' or ``flux-frozen'' accretion disks that form in the simulations. As we explain the model predictions below, we will compare them to the simulations from \papertwo\ in Fig.~\ref{fig:sim.model.comparison},\footnote{Note that the specific time plotted and weighting of the mean for the simulations is different in Fig.~\ref{fig:sim.model.comparison} from similar plots in \papertwo. This is because we focus on volume and time-averaged quantities in the midplane here, which can be most directly compared to the models (which do not include e.g.\ effects of small-scale multi-phase structure in the outer disk, which were discussed in \paperone\ and \papertwo).} over the range where one can reasonably identify a rotationally-dominated ``accretion disk'' in such simulations ($r \sim (300-10^{6})\,2\,G\,M_{\rm BH}/c^{2}$).\footnote{The inner boundary of the simulation is at $\approx 80\,{\rm au}\approx 4\times 10^{-4}\,{\rm pc}$. The actual outer simulation boundary is at $\sim 100\,$Mpc (recall this is zoomed-in from a fully cosmological box), but the outer boundary at which any sort of ``accretion flow'' can be meaningfully defined or the models here defined is $R_{\rm BHROI} \approx 5\,$pc in the simulation. So we plot in-between, excluding a factor of a couple on either side to avoid boundary effects (but if we plot the entire range from $0.0004-5\,$pc, our model behaves similarly well). Consistent with our model, the outermost radii where there is a clearly circularized, rotation dominated ``disk'' are more like $\sim 1\,$pc (see \papertwo), while $\sim 2-5\,$pc resembles more of a ``free-fall zone.''} We stress that the model predictions are not fitted to the simulations -- we simply adopt the boundary conditions (for values like $\dot{M}$ and $M_{\rm bh}$) defined therein.

\subsection{Solutions for the Accretion, Magnetic Field, and Disk Structure}
\label{sec:analytic.structure}

Denote the thermal sound, \Alf, circular ($V_{c} \equiv \sqrt{G M_{\rm enc}(<r)/r}$ in terms of total enclosed mass $M_{\rm enc}$), and effective ``turbulent'' speeds as $c_{s}$, $v_{A}$, $V_{c}$, $v_{t}$, respectively. Motivated by the simulations, we will assume a disk geometry, and that the disk is magnetic-pressure-dominated, super-sonically and trans-\Alf{ically} turbulent, efficiently cooling ($t_{\rm cool} \lesssim t_{\rm dyn}$), and features angular momentum transfer dominated by some combination of Maxwell and Reynolds stresses (discussed below). We will neglect angular momentum transport via magnetically-driven outflows (these are negligible in the simulations). We validate the self-consistency of all these assumptions below. For simplicity, we focus on similarity solutions and denote the relevant $\mathcal{O}(1)$ dimensionless pre-factors with $\varpi_{X} \sim 1$ for different quantities $X$. For generality, we will write our scalings in terms of some arbitrary orbital frequency $\Omega \equiv V_{\rm c}/r$, but for the regime of greatest interest here we can safely assume the Keplerian limit $\Omega \rightarrow  v_{\rm K}/r = \sqrt{G M_{\rm bh}/r^{3}}$ (i.e.\ $V_{c} \approx v_{\rm K} = \sqrt{G M_{\rm bh}/r}$).

From these assumptions, we expect an effective viscosity\footnote{Here, we introduce $v_{e}^{2}$ which can have contributions $\mathcal{O}(v_{t}^{2})$ from the Reynolds and (especially for primarily anti-correlated toroidal-radial fields as seen in the simulations) $\mathcal{O}(v_{A}^{2})$ from the mean and/or fluctuating Maxwell stresses. These are comparable in the simulations and would appear with a fixed ratio here, so we can take $v_{e}^{2} \sim v_{t}^{2}$ with their relative normalization and the appropriate $\mathcal{O}(1)$ prefactors subsumed into $\varpi_{\nu}$. As noted below, this means that if the mean Maxwell $B_{\phi}-B_{R}$ stress has the correct properties, the disks do not strictly need to be strongly turbulent, though they appear to be in practice.} $\nu_{\rm visc} \approx \varpi_{\nu}\,v_{e}^{2}/\Omega$ giving rise to an inflow rate $\dot{M} = 3\pi\,\nu_{\rm visc}\,\Sigma_{\rm gas}$. Equivalently, the energy lost to turbulent dissipation, which is ultimately radiated away, is balanced by gravitational energy from accretion: 
\begin{align}
\label{eqn:eqm.disk} \frac{d E_{\rm turb+mag}}{d t\,dA} &\approx \frac{\varpi_{t}}{2}\,v_{t}^{2}\,\Sigma_{\rm gas}\,\Omega \approx  \frac{d W_{\rm grav}}{d A} \approx \varpi_{g}\,\frac{3}{4\pi}\,\dot{M}\,\Omega^{2}
\end{align}
so by definition $9\,\varpi_{\nu}\,\varpi_{g} \approx 2\,\varpi_{t}$ (and recall $\varpi_{t} \sim \varpi_{g} \sim \varpi_{\nu} \sim 1$; for more detailed discussion see \citealt{pariev:2003.mag.dominated.disk.models}). 
Since the disks are magnetically dominated, the disk scale-height $H \approx \varpi_{H}\,v_{A}/\Omega \equiv \varpi_{H}\, B/\Omega\sqrt{4\pi\,\rho_{\rm mid}}$ ($\varpi_{H} \sim 1$), in terms of the midplane density $\rho_{\rm mid} \equiv \varpi_{\rho}\,\Sigma_{\rm gas}/2\,H$, so (re-arranging) we have $H \approx (\varpi_{H}\,B/\Omega)^{2}/(2\pi\,\varpi_{\rho}\,\Sigma_{\rm gas})$. From Eq.~\eqref{eqn:eqm.disk} we can re-arrange to write the gas surface density as $\Sigma_{\rm gas} \approx \dot{M}\,\Omega/(3\pi\,\varpi_{\nu}\,v_{t}^{2})$. 

In \papertwo, it is shown that the magnetic-field strengths in the simulations arise from simple flux-freezing/flux-advection considerations, and indeed even the simplest-possible flux-freezing argument, which predicts $|{\bf B}| \propto \rho^{2/3}$ in the midplane of the disk, works quite well at explaining the mean trend of $|{\bf B}(\rho)|$ and $|{\bf B}(r)|$. We will therefore assume such a scaling here (though we discuss below the weak effects of adopting some different prescription for ${\bf B}(\rho)$). Next, since we are interested in similarity solutions, we parameterize $v_{t} \propto r^{-\zeta}$. Comparison with the simulations in \papertwo\ and basic physical considerations suggest a plausible range of $0\lesssim \zeta \lesssim 1/2$, for which we will see that the solutions give a radial inflow velocity $-v_{r}/v_{\rm K} \approx \varpi_{\nu}\,(v_{t}/v_{\rm K})^{2}$ which increases with $r$. Therefore, it is convenient to normalize our solutions to an outer radius, which we call the ``freefall radius'' $r_{\rm ff}$, at which point $v_{r}(r_{\rm ff}) \approx -v_{\rm K}$; i.e.\ at $r\sim r_{\rm ff}$ the gas is in near free-fall (at the Keplerian speed) onto the center. The solutions we derive therefore patch onto a solution of gas free-falling onto the nuclear region at $r \gtrsim r_{\rm ff}$, as in the simulations. We can now write: 
\begin{align}
\label{eqn:disk.general.profile} \frac{v_{t}}{v_{\rm K}} &\approx \sqrt{\frac{2}{3\,\varpi_{\nu}}}\,\tilde{r}^{\frac{1}{2}-\zeta} \\ 
\frac{v_{A}}{v_{\rm K}} &\approx \psi_{A}\,\tilde{r}^{\frac{1}{14}+\frac{2\,\zeta}{7}} \\ 
\Sigma_{\rm gas} &\approx \frac{\dot{M}}{2\pi\,\sqrt{G\,M_{\rm bh}\,r_{\rm ff}}}\,\tilde{r}^{-\frac{3}{2}+2\,\zeta}
\end{align}
where $\tilde{r} \equiv r/r_{\rm ff}$, and $\psi_{A} \equiv v_{A}/|v_{\rm r}|$ at $r_{\rm ff}$ provides the outer boundary condition for our flux-freezing assumption, normalizing $B$. 

Motivated by the results in \papertwo, let us further assume that the turbulence/stresses are approximately trans-\Alf{ic}, being set by an initially turbulent field that is not order-of-magnitude different from isotropic, so $v_{t} \sim v_{A}$. This is equivalent to taking $\psi_{A} \rightarrow (2/3\varpi_{\nu})^{1/2} \sim 1$ and $\zeta \rightarrow 1/3$, giving,
\begin{align}
\label{eqn:veff.model}\frac{v_{t}}{v_{\rm K}} &\sim \frac{v_{A}}{v_{\rm K}} \sim \frac{H}{R} \sim \tilde{r}^{1/6} \sim 0.07\,\left( \frac{m_{7}\,x_{g}}{r_{{\rm ff},5}} \right)^{1/6}, \\
\label{eqn:sigma.gas} \frac{\Sigma_{\rm gas}}{{\rm g\,cm^{-2}}} &\sim 0.01 \frac{\dot{m}\,m_{7}^{1/2}}{r_{{\rm ff},5}^{1/2}\,\tilde{r}^{5/6}} \sim 10^{4} \frac{\dot{m}\,r_{{\rm ff},5}^{1/3}}{m_{7}^{1/3}\,x_{g}^{5/6}},
\end{align}
where we define $\dot{m} \equiv \dot{M}/\dot{M}_{\rm edd} \equiv t_{\rm S}\,\dot{M}/M_{\rm bh}$ as the accretion rate in Eddington units (with a Salpeter time $t_{\rm S} \equiv 5\times10^{7}\,$yr defined for a reference radiative efficiency $\approx 0.1$), $m_{7} \equiv M_{\rm bh}/10^{7}\,{\rm M_{\odot}}$, $r_{{\rm ff},5} \equiv r_{\rm ff} / (5\,{\rm pc})$.  The second scaling in each expression normalizes $r$ to the gravitational radius of the BH, $x_{g} \equiv r/R_{g}$ with $R_{g} \equiv G\,M_{\rm bh}/c^{2}$, instead of normalizing $r$ to $r_{\rm ff}$.\footnote{If one wishes to scale to e.g.\ accreting binaries and stellar-mass BHs, it is helpful to note $m_{7}/r_{{\rm ff},5} \approx 25\,(m_{1}/P_{{\rm ff},{\rm day}})^{2/3}$ where $m_{1}=m/10\,{\rm M_{\odot}}$ and $P_{{\rm ff},\,{\rm day}}$ is the orbital period at $r_{\rm ff}$ in days. So Eqs.~\eqref{eqn:veff.model}-\eqref{eqn:sigma.gas} become $H/R\sim 0.1\,(m_{1}/P_{{\rm ff},{\rm day}})^{1/9}\,x_{g}^{1/6}$ and $\Sigma_{\rm gas} \sim 3000\,(P_{{\rm ff},{\rm day}}/m_{1})^{2/9}\,\dot{m}\,x_{g}^{-5/6}$, and Eq.~\eqref{eqn:beta} becomes $\beta \sim 3\times10^{-4}\,(\dot{m}/f_{\tau})^{1/4}\,x_{g}^{-1/12}\,P_{{\rm ff},{\rm day}}^{2/9}\,m_{1}^{-17/36}$.}

We can immediately check the consistency of our assumptions: $v_{t} \sim v_{A} \lesssim v_{\rm K}$ at all $r < r_{\rm ff}$ (the regime of validity here), so the disk is thin/slim ($H < r$) and the accretion will not be magnetically ``arrested''; the turbulence is trans-\Alf{ic} (by assumption); even at the ISCO ($x_{g} \sim 6$), $v_{A}/c \sim 0.04\,(m_{7}/r_{{\rm ff},5})^{1/6} \sim 0.04\,m_{7}^{1/12}$ so for all interesting BH masses, it is reasonable to neglect relativistic corrections at this order; the total effective (magnetic + turbulent + thermal) Toomre $Q$ parameter $Q_{\rm tot} = \sigma_{\rm eff}\,\kappa/\pi\,G\,\Sigma_{\rm gas}$ (with $\sigma_{\rm eff}^{2} \approx v_{t}^{2} + v_{A}^{2} + c_{s}^{2}$, and $c_{s}$ small by assumption) is:
\begin{align}
\label{eqn:Q} Q_{\rm tot} &\approx 3000\frac{m_{7}^{1/2}}{r_{{\rm ff},5}^{3/2}\,\dot{m}\,\tilde{r}} \sim \frac{3000}{m_{7}^{1/4}\,\dot{m}\,\tilde{r}}
\end{align}
so the disk is stable at all radii $r < r_{\rm ff}$ (outside of $r > r_{\rm ff}$, ISM physics applies) so long as $\dot{m} < 3000\,m_{7}^{-1/4}$ -- i.e.\ even for highly super-Eddington accretion rates. 

We can also confirm that various more subtle properties observed in the simulations should hold for this model. Per the arguments in \papertwo\ (\S~6.6 therein), if we assume the field $|{\bf B}|$ is initially dominated by some combination of radial/toroidal fields at $\sim r_{\rm ff}$ (as seen in the simulations, and expected for free-falling/circularizing material tidally captured a turbulent ISM beyond the BHROI), then we would have fields with inverse coherence length $k_{B_{R,\,\phi}} \sim 1/R$ at $r\sim r_{\rm ff}$. Given this, the model suggests that (1)  the radial and toroidal fields will  maintain $k_{B_{R,\,\phi}} \sim 1/R$ as the flow migrates inwards (i.e.\ the field lines will be ``stretched'' inwards in the $R$, $\phi$ plane, requiring $\partial\ln{(-v_{R}/R)}/\partial{\ln{R}} > 0$) so long as $\zeta \gtrsim -1/4$ (true for all solutions we consider); (2) the supply of $B_{\phi}$ from advection of radial flux (${\rm d}_{t} B_{\phi}  \sim \partial_{R}  v_{\phi}\,B_{R} +...$) will maintain the $B_{\phi}-B_{R}$ anti-correlation needed to ensure the Maxwell stress transports angular momentum outwards (\papertwo, \S~5); and (3) the turbulent resistivity, ${\rm d}_{t}^{\rm turb} B_{R} \sim -\eta_{\rm turb} \,k_{B_{R}}^{2}\,B_{R} \sim -\tilde{\eta}\,(v_{t}/v_{\rm K})^{2}\,\Omega\,B_{R}$ with $\tilde{\eta} \lesssim 1$, will not  damp the fields signficantly faster than they are advected inwards, since $|{\rm d}_{t}^{\rm turb}|$ cannot be $\gg|{\rm d}_{t}^{\rm advection}|\sim k_{B_{R}} v_r\sim \varpi_\nu (v_{t}/v_{\rm K})^{2}\,\Omega$ (though they can  be of similar magnitudes). For a much more detailed discussion and justification of these points, see \papertwo. 

Comparing to the simulations in Fig.~\ref{fig:sim.model.comparison} (taking the simulation values of $M_{\rm bh}$, $\dot{M}$, and $r_{\rm ff} \sim R_{\rm BHROI}$ directly, so there is no ``adjustable parameter'' to be fit here), the model Eqs.~\ref{eqn:veff.model}-\ref{eqn:sigma.gas} appears to describe the simulations remarkably well (at the order-of-magnitude level) and capture most of the important qualitative scalings (though of course there are obvious deviations from strictly self-similar, pure-power-law, steady-state behavior in the simulations). For comparison, the traditional \citet{shakurasunyaev73} $\alpha$-disk model contrasted in Fig.~\ref{fig:sim.model.comparison} differs from the simulations by as much as factors of $\sim 10^{8}$ in key properties such as the Toomre $Q$ or $\beta$ or $\Sigma_{\rm gas}$.

\subsection{Thermal \&\ Radiation Structure}
\label{sec:analytic.thermal}

Note that because the disk is magnetically-dominated we did not need to invoke assumptions about its thermal properties to solve for its structure. However it is useful to do so for its own sake and to validate the consistency of our assumptions. Begin with the usual optically thick, thermal disk assumption: $T\approx T_{\rm gas,\,mid} \approx T_{\rm rad,\,mid}$ with cooling flux $d L_{\rm cool}/dA \sim 2\,\sigma_{B}\,f_{\tau}\,T^{4}$ (where $f_{\tau}$ is some function that accounts for optical depth effects, vertical stratification, etc.). Taking $d L_{\rm cool}/dA \sim dE_{\rm turb+mag}/dt\,dA$ we can solve for $T$ or $c_{s} \equiv \sqrt{k_{B}\,T/\mu\,m_{p}}$ and obtain,
\begin{align}
\beta &\equiv \frac{c_{s}^{2}}{v_{A}^{2}} \approx 10^{-5}\,\frac{\dot{m}^{1/4}\,r_{{\rm ff},5}^{1/4}\,\tilde{r}^{3/28-4\zeta/7}}{\psi_{A}^{2}\,f_{\tau}^{1/4}\,m_{7}^{1/2}}.
\end{align}
We can also solve for the ratio of cooling to dynamical or free-fall time $t_{\rm cool}/t_{\rm dyn} \sim (3/4)\,(\Sigma_{\rm gas}\,c_{s}^{2}\,\Omega)/(\sigma_{B}\,f_{\tau}\,T^{4}) \sim (c_{s}/v_{t})^{2}$, sonic Mach number $\mathcal{M}_{s} \sim v_{t}/c_{s}$, or ``equivalent Shakura-Sunyaev $\alpha$-parameter'' $\alpha_{\rm therm,\,equiv}$ defined below, which for $\zeta \sim 1/3$ gives: 
\begin{align}
\nonumber \beta &\sim \frac{t_{\rm cool}}{t_{\rm dyn}} \sim \frac{1}{\mathcal{M}_{s}^{2}} \sim \frac{1}{\alpha_{\rm therm,\,equiv}} \sim 10^{-5}\,\frac{\dot{m}^{1/4}\,r_{{\rm ff},5}^{1/4}}{f_{\tau}^{1/4}\,m_{7}^{1/2}\,\tilde{r}^{1/12}} \\
\label{eqn:beta} & \sim 3\times 10^{-5}\frac{\dot{m}^{1/4}\,r_{{\rm ff},5}^{1/3}}{f_{\tau}^{1/4}\,m_{7}^{7/12}\,x_{g}^{1/12}}.
\end{align}

We can immediately validate several assumptions: for any reasonable $f_{\tau}$, $\beta \ll 1$ is very small and depends  weakly on radius so this is true at all radii from the freefall radius to the ISCO.\footnote{If we used $f_{\tau}^{-1} = 1+\tau^{-1} + 3\,\tau/8$ derived for a stratified LTE disk with $\tau \equiv \kappa_{R}(T=T_{\rm mid})\,\Sigma_{\rm gas} \lesssim 1$, then this does not change our conclusions regarding $\beta \ll 1$. For the outermost disk where $\tau\lesssim 1$ is possible, this generically gives $\beta$ {\em decreasing} towards smaller $r$ from $r_{\rm ff}$ (closer to what we actually see in the simulations at larger radii in \paperone). For $\tau \gg 1$, if H$^{-}$ opacity dominates (as in the outer disk in \papertwo), its strong $T^{7.7}$ temperature dependence means again $\beta$ decreases at smaller $r$ from $r_{\rm ff}$; at smaller $r$, if Kramers and/or electron scattering opacity dominate and $\tau \gg 1$ for this $f_{\tau}$, it only slightly changes our coefficients (giving e.g.\ $\beta \propto \tilde{r}^{-1/9}$ instead of $\tilde{r}^{-1/12}$, not enough to change our conclusions).}
 Correspondingly, the turbulence is highly super-sonic ($\mathcal{M}_{s} \gg 1$), and balanced by rapid cooling with $t_{\rm cool} \ll t_{\rm dyn}$. We can also calculate the ``thermal-only'' Toomre $Q_{\rm therm} \equiv c_{s}\,\kappa/\pi\,G\,\Sigma_{\rm gas}$, which gives $Q_{\rm therm} \sim 6\,m_{7}^{1/4}\,f_{\tau}^{-1/8}\,\dot{m}^{-7/8}\,r_{{\rm ff},5}^{-11/8}\,\tilde{r}^{-25/24}$, so despite $c_{s} \ll v_{A}$ the thermal $Q_{\rm therm} \gg 1$ at most radii (and even at extremely large radii of several pc, it only falls to unity or lower for $\dot{m} \gg 10\,m_{7}^{-1/2}$, which is consistent with the behaviors seen and discussed at radii $>$\,pc in \paperone). This is important in that it means the disk is only very weakly fragmenting even where e.g.\ collapse along field lines in magnetic field switches is possible, and becomes fully stable against e.g.\ gravito-turbulence at radii $\lesssim$\,pc (for SMBH masses), as discussed and shown in \paperone\ \citep[see also][]{pariev:2003.mag.dominated.disk.models}. 

At large radii, this scaling for $T$ will break down (as we note below), but even if we assume the temperature there is determined by more complicated (NLTE, non-equilibrium, atomic+molecular) ISM thermo-chemistry, and set the temperature to some isothermal $T\sim 10^{3}-10^{4}\,$K (a relatively high value for the largest radii, see \paperone), we obtain $\beta \lesssim 0.001\,\tilde{r}^{2/3}\,m_{7}^{-1/2}$ and still have $\beta \sim t_{\rm cool}/t_{\rm dyn} \sim 1/\mathcal{M}_{s} \sim 1/\alpha_{\rm therm,\,equiv}$. Thus, the basic assumptions of the model remain safe (and now $Q_{\rm therm} \sim (60-200)\,\dot{m}^{-1}\,r_{{\rm ff},5}^{-1}\,\tilde{r}^{-2/3}$ is larger, as expected).  

Similarly, computing the ratio of radiation pressure ($P_{\rm rad} = a\,T_{\rm mid}^{4}/3$) to magnetic+turbulent pressure gives: 
$P_{\rm rad}/P_{\rm other} \sim (0.1/f_{\tau})\,(m_{7}/r_{{\rm ff},5})^{1/6}\,x_{g}^{-1/3} \sim (0.0006/f_{\tau})\,(m_{7}/r_{{\rm ff},5})^{1/2}\,\tilde{r}^{-1/3}$. This is safely $\ll 1$ at larger radii of order $r_{\rm ff}$ (even more so if the disk is optically thin and/or atomic). If $f_{\tau} \sim 1$ in the inner regions of the disk (and using $r_{{\rm ff},5}\sim m_{7}^{1/2}$ below) then this is safely $\ll 1$ even at the ISCO ($x_{g} \sim 6$), but we discuss this limit in more detail below.

In the inner disk, to understand $f_{\tau}$ and the vertical stratification (or lack thereof), first let us estimate $\tau$. Scaling to the electron scattering opacity $\kappa_{\rm es}$ gives $\tau \sim 3000\,\dot{m}\,(r_{{\rm ff},5}/m_{7})^{1/3}\,x_{g}^{-5/6}\,(\kappa/\kappa_{\rm es})$, and estimating the absorption opacity as the Kramers opacity gives that the optical depth to thermalization is $\gtrsim 1$ where $\tau \gg 1$. However, comparing the vertical mixing/transport timescale via turbulence, $t_{\rm mix} \sim H/v_{t} \sim \Omega^{-1}$ from the above, to the radiation diffusion timescale required to establish vertical thermal stratification $t_{\rm diff} \sim \tau\,H/c$, gives $t_{\rm mix}/t_{\rm diff} \sim 0.004\,\dot{m}^{-1}\,(m_{7}/r_{{\rm ff},5})^{1/6}\,x_{g}^{7/6}\,(\kappa_{\rm es}/\kappa) \ll 1$. Thus we expect efficient vertical mixing to make the disk weakly thermally stratified (meaning that we do not expect a strong temperature gradient at $|z| \ll H$), leading to $f_{\tau} \sim 1$. We discuss this case further in \S~\ref{sec:analytic.eddington}. Given these ratios, it also follows that the inner disk is stable against the usual viscous and thermal instabilities, per the discussion in \citet{begelman.pringle:2007.acc.disks.strong.toroidal.fields} \S~4.1.3. 

In the outer disk, the surface densities are low and two things happen: first, from Eq.~\ref{eqn:sigma.gas} we see as $\tilde{r} \rightarrow 1$, $\Sigma_{\rm gas}$ can be quite small and so even assuming full ionization with electron scattering opacities, one finds $\tau \lesssim 1$ (let alone the absorption/Kramers opacity, which will be smaller). Moreover, the equilibrium temperature assuming blackbody-like cooling becomes quite low, $T_{\rm BB} \rightarrow 10\,{\rm K}\,\tilde{r}^{-3/4}\,\dot{m}^{1/4}\,m_{7}^{1/8}$, so the gas becomes primarily atomic at radii $\tilde{r} \gtrsim (0.2-1) \times 10^{-4}\,m_{7}^{1/6}\,\dot{m}^{1/3}$ -- just as we see in our simulations (see \paperone). So the opacity drops significantly and the disk can even become multi-phase. At these radii the thermal physics can be influenced by optical depth effects, but are set by a  more complicated balance of heating from external radiation (stars in the galaxy and the CMB, which at typical quasar redshifts may not be negligible), cosmic rays, thermal cooling, non-equilibrium chemistry, etc. These maintain a weakly-stratified, cool disk with $\beta \ll 1$ per the discussion above (with warm atomic gas from a few hundred to $\sim 8000\,$K), and $P_{\rm rad}\, , \, P_{\rm therm} \ll P_{\rm mag}$.

Finally, it is easy to verify from the temperatures and chemical structure above that the microphysical resistivity and non-ideal MHD effects (e.g.\ ambipolar diffusion, the Hall effect, Ohmic resistivity) are negligible for all disk conditions of interest here, as expected (and verified in the simulations directly; see \papertwo).

\section{Sensitivity to Assumptions and Caveats}
\label{sec:analytic.caveats}

It should be obvious that the scalings above represent a tremendously simplified model. We can see in Fig.~\ref{fig:sim.model.comparison} that in the ``full physics'' simulations, even over a modest dynamic range, are not precisely similarity solutions (the ``effective power law index'' of different quantities changes with scale); nor  are they in exact steady state (power-law indices vary with time, and quantities like $\dot{M}$ are not exactly constant with $r$ as required in a strictly steady-state model; see \papertwo); nor, as discussed in \papertwo, is the disk exactly  axisymmetric (it has large coherent eccentricity and smaller-scale spiral modes). But in Fig.~\ref{fig:sim.model.comparison}, we plot the analytic expectations from this model and show that they nonetheless do appear to capture the key qualitative behaviors and at least order-of-magnitude values of the most important quantities in the simulations. 

We can also vary some of the detailed assumptions within our analytic model. From examination of the simulations we see $\zeta$ is not exactly constant in space or time and can vary between $0 \lesssim \zeta \lesssim 1/2$. Varying over this range gives slightly different results but does not change any of our conclusions: for example, we still obtain $\beta \ll 1$ with $\beta \propto r^{q}$ with a very weak dependence on radius $-5/28 < q < 3/28$. We can vary $\psi_{A}$ but so long as it is not extremely small ($\psi_{A}\ll 0.01$), which would be inconsistent with the simulations motivating this model in the first place,  we still obtain similar disk structure and $\beta \ll 1$, $Q \gg 1$. We have also considered parameterizations of the form $v_{t} \propto v_{A}^{\eta}\,v_{\rm K}^{1-\eta}$ for some $\eta$, but this is just equivalent to a narrower range of $1/3<\zeta<1/2$ and $\psi_{A} \sim 1$. We have also re-calculated the entire model assuming the alternative flux-freezing relation discussed in \papertwo,  $B_{\phi} \propto H^{-1}$ (which is also broadly consistent with the simulations). For a given $\zeta$, assuming $B_{\phi} \propto H^{-1}$ gives us identical power-law scaling for $v_{t}/v_{\rm K}$ and $\Sigma_{\rm gas} \propto \tilde{r}^{-3/2+2\,\zeta}$, with $\beta \propto \tilde{r}^{-3/4+4\,\zeta/3}$, and for e.g.\ $v_{A}/v_{\rm K} \sim H/R$ modifies the $\tilde{r}$ dependence to $\tilde{r}^{1/2-2\,\zeta/3}$. Comparing this and/or $\Sigma_{\rm gas}$ to the simulations would, with this model, favor something more like $\zeta\sim 3/8$, modifying $v_{A}/v_{\rm K}$ to be $\propto \tilde{r}^{1/4}$ instead of $\propto \tilde{r}^{1/6}$ (and $\Sigma_{\rm gas} \propto \tilde{r}^{-3/4}$, $v_{t}/v_{A} \propto \tilde{r}^{-1/8}$), all fairly weak effects.\footnote{We have also re-computed the model using the flux scaling assumed in the analytic models of \citet{oda:2009.analytic.mag.disk.structure.models}  $B_{\phi} \propto 1/(H\,|v_{r}|)$ (although this agrees less well with our simulations), which gives which gives $v_{A}/v_{\rm K} \propto 1/6 + 2\zeta/3$, so a value $\zeta \sim 1/5$ provides a reasonable match to the simulations with $\Sigma_{\rm gas} \propto r^{-11/10}$, 
$v_{t}/v_{A} \sim $\,constant.} But even for this closure and more extreme assumptions like $\zeta\sim 0$ or $B_{\phi} \propto r^{-1}$, which would introduce a stronger (though still sub-linear) dependence of $v_{A}/v_{\rm K}$ on $\tilde{r}$ (and deviate more notably from the simulation behaviors which motivate these models), this would not change our qualitative conclusions at large radii, but would extrapolate inwards differently and would lead to a thinner, more thermal-dominated disk at small $r$ (approaching the inner disk and ISCO). This could produce a limit more like that discussed in \citet{begelman.pringle:2007.acc.disks.strong.toroidal.fields} and \S~\ref{sec:analytic.bp07} below in the inner disk (with the parameter $B_{\phi}/B_{\rm BP07}$ defined in \S~\ref{sec:analytic.bp07} falling to $\lesssim 1$ as $\tilde{r} \rightarrow 0$).

More generally, for power-law solutions like those here, we can consider the generalized ``closure'' relation $P_{B} \propto B^{2} \propto \rho^{\gamma}$ (our default model takes $\gamma=4/3$), and note this is equivalent to any arbitrary power-law function $|B| \propto r^{\gamma^{\prime}_{r}}\,H^{\gamma^{\prime}_{H}}\,\Sigma^{\gamma^{\prime}_{\Sigma}}\,v_{r}^{\gamma^{\prime}_{v}}\,...$ for appropriate $\gamma$ (for example, $B^{2}\propto \rho^{\gamma}$ gives $B\propto r^{\gamma^{\prime}_{r}}$ with $\gamma^{\prime}_{r} = (2\,\gamma\,\zeta - 3\gamma)/(1+\gamma)$). In Eq.~\eqref{eqn:disk.general.profile}, this modifies $v_{A}/v_{\rm K} \rightarrow \psi_{A} \tilde{r}^{(7-5\gamma-4\zeta+4\gamma\zeta)/(2+2\gamma)}$ (while $v_{t}/v_{\rm K}$ and $\Sigma_{\rm gas}$ scale identically). Per the discussion above, if we still assume trans-\Alf{ic} turbulence (and/or a broadly similar range of $\zeta$), the model predictions are quantitatively modified but remain qualitatively similar over a wide dynamic range in $\tilde{r}$ for any closure in the plausible range $1\lesssim \gamma \lesssim 2$. For $\gamma \gtrsim 2$ (depending on the exact value of $\zeta$ and the boundary condition $\psi_{A}$), $v_{A}/v_{\rm K}$ and $H/R$ will rise sufficiently steeply as $r\rightarrow 0$ such that they could exceed unity before reaching the ISCO -- this implies the disk would transition to a more MAD-like state at small radii (unless $\zeta \gtrsim 1/2$, in which case they would instead transition back to free-fall solutions). For $\gamma \lesssim 0.8$ or so, such that the \Alf\ speed $v_{A} \propto \rho^{(\gamma-1)/2}$ {\em decreases} at higher densities and smaller radii within the disk, $v_{A}/v_{\rm K}$  will fall sufficiently rapidly as $r\rightarrow 0$ so that $\beta$ increases and exceeds unity before reaching the ISCO. This would imply a transition either to a SS73-like disk at small-$r$ or perhaps an intermediate disk more akin to the \citet{begelman.pringle:2007.acc.disks.strong.toroidal.fields} model discussed in  \S~\ref{sec:analytic.bp07}.

Briefly, it is worth noting that a particularly simple and interesting model variant arises if we assume $\gamma=5/3$ and trans-\Alf{ic} turbulence, which gives: 
\begin{align}
\label{eqn:veff.stiff} \frac{v_{t}}{v_{\rm K}} &\approx \frac{v_{A}}{v_{\rm K}} \approx \frac{H}{R} \approx \psi_{A} = {\rm constant} \ , \\ 
\label{eqn:sigma.stiff} \Sigma_{\rm gas} &\approx \frac{\dot{M}}{2\pi\,\psi_{A}^{2}\sqrt{G\,M_{\rm bh}\,r}} \ , \\ 
\label{eqn:Q.stiff} Q_{\rm tot} &\approx 2^{3/2}\,\psi_{A}^{3}\,\frac{\Omega\,M_{\rm bh}}{\dot{M}}\ .
\end{align}
While not quite as good a fit in Fig.~\ref{fig:sim.model.comparison} (particularly at $R \gtrsim 0.01\,$pc), in this case the flow is truly self-similar, is entirely parameterized by one dimensionless $\mathcal{O}(1)$ number $\psi_{A}$ (note $r_{\rm ff}$ entirely factors out of the solutions), and maps  onto free-fall with $v_{r} \sim c$ approaching the horizon. Given that the disk mass $M_{\rm gas}(<r) < M_{\rm bh}$ within the BHROI (by definition), and that accretion cannot be faster than dynamical ($\dot{M} \lesssim \psi_{A}\,M_{\rm gas}(R_{\rm BHROI})\,\Omega(R_{\rm BHROI})$), it becomes basically impossible to have $Q < 1$ for any physical inflow rate at the outer boundary (the upper limit is set by supply from galactic scales). Also here $\beta \sim t_{\rm cool}/t_{\rm dyn} \sim 1/\mathcal{M}_{s}^{2} \sim 1/\alpha_{\rm therm,\,equiv} \sim 10^{-5}\,\psi_{A}^{-2}\,(\dot{m}\,r_{5}/f_{\tau}\,m_{7}^{2})^{1/4}$ ($r_{5} \equiv r/{\rm 5\,pc}$) is similar to our default model in the outer disk and decreases weakly (for reasonable $f_{\tau}$) as $r\rightarrow 0$, so the disk becomes more magnetically-dominated. All of our qualitative conclusions regarding consistency and relative importance of different terms for our ``default'' parameterization apply to this solution as well.

Per the discussion in \S~\ref{sec:analytic.thermal} and Fig.~\ref{fig:sim.model.comparison}, the least robust part of our analytic model in general (comparing to the simulations) is probably the description of the thermal disk properties at the largest radii, although as noted there, this has no effect on the predicted accretion rates or mass profile since $\beta \ll 1$. This owes primarily to the facts that LTE is a poor approximation, the disk is highly super-sonically turbulent, and (at large radii) the optical depths can become modest and the disk is no longer mostly ionized. More detailed thermochemical modeling is therefore warranted.

\subsection{How Generic vs.\ Fine-Tuned Are the Parameters?}
\label{sec:analytic.general}

Given the discussion above (\S~\ref{sec:analytic.caveats}), the qualitative behaviors in the model are quite robust to details of e.g.\ the detailed profile slopes ($\zeta$), scaling of $v_{t}$, etc., and as argued in \papertwo\ the key dynamics and magnetic structure of the disk are set by simple flux-freezing/advection considerations and essentially independent of the detailed thermal structure. If those are general statements, then so long as the outer boundary conditions -- encapsulated in our analytic model above by two parameters $r_{\rm ff}$ and $\psi_{A}$ -- have ``reasonable'' values for a given $M_{\rm bh}$ and $\dot{M}$, and are not especially ``fine-tuned,'' it suggests this scenario may be robust. The ``fine-tuned'' question can be addressed immediately, as the dependence of key properties ($H/R$, $v_{t}/v_{A}$, etc.) on $r_{\rm ff}$ and $\psi_{A}$ is quite weak (see Eq.~\ref{eqn:disk.general.profile}-\ref{eqn:sigma.gas}), so no fine-tuning is required.

Regarding ``reasonable'' values, recall $r_{\rm ff}$ represents the radius at which the solution maps onto approximate free-fall onto the BH. It is notable, then, that our reference value of $r_{\rm ff}$, taken from the simulation, is very similar to the BHROI $R_{\rm BHROI} \equiv G\,M_{\rm bh}/\sigma_{\rm gal}^{2}$ where $\sigma_{\rm gal}$ is the velocity dispersion of the galaxy center/nucleus/bulge region. Indeed this is exactly what we expect (order-of-magnitude) for $r_{\rm ff}$ if the accretion is ultimately driven by gravitational capture of gas from ambient complexes outside the BHROI moving at a characteristic velocity $\sim \sigma_{\rm gal}$ (precisely what is seen in the simulations, see \paperone). If we further assume that BHs lie on the observed galactic $M_{\rm bh}-\sigma_{\rm gal}$ relation \citep{kormendy:2013.review.smbh.host.correlations}, we have $r_{{\rm ff},5} \sim m_{7}^{1/2}$, which we can insert in the scalings above to eliminate $r_{\rm ff}$, making the dependence on $M_{\rm bh}$ even weaker.

Next, consider $\psi_{A}$, which normalizes the magnetic field strength. For any system where the inflow velocities are trans or super-\Alf{ic} on large (galactic) scales (outside the BHROI), $\psi_{A} \sim 1$ is a natural expectation. Moreover, recall we have only assumed flux-freezing, with $B \propto \rho^{2/3}$, so $\psi_{A}$ is equivalent to the normalization of this relation. As demonstrated explicitly in \papertwo, for the parameters in these simulations (these values of $M_{\rm bh}$, $\dot{M}$, and $r_{\rm ff}$), $\psi_{A} \sim 1$ is equivalent to $B \sim 8\,{\rm \mu G}\,(n / {\rm cm^{-3}})^{2/3}$ -- i.e.\ completely typical ISM values \citep{crutcher:cloud.b.fields,ponnada:fire.magnetic.fields.vs.obs}. More generally, for arbitrary $\zeta$ and $\psi_{A}$ if we assume $B\propto \rho^{2/3} \approx b_{10}\,10\,{\rm \mu G}\,(n/{\rm cm^{-3}})^{2/3}$ strictly extrapolated from the ISM, we have $\psi_{A} \sim (2/3)\,\dot{m}^{1/7}\,b_{10}^{6/7}\,r_{\rm ff,\,5}^{3/14}\,m_{7}^{-5/14}$ ($\sim 0.7\,\dot{m}^{1/7}\,m_{7}^{-1/4}\,b_{10}^{6/7}$ if we assume $r_{\rm ff}$ traces the BHROI, or $\sim 1.8\,m_{7}^{-2/7}\,b_{10}^{6/7}$ if we assume $\dot{m}$ traces free-fall of an enclosed gas mass $\sim M_{\rm bh}$ at $r_{\rm ff}$). In other words, even if there were no amplification mechanism ``driving'' the system towards trans-\Alf{ic} turbulence (though in \paperone\ and \papertwo\ we show there are multiple such mechanisms), we would still expect $\psi_{A} \sim 1$ given the observed scalings of ISM magnetic field strengths. And to qualitatively change our conclusions regarding e.g.\ $\beta \ll 1$ or $Q$, we would need to lower the extrapolated $B$ (e.g.\ $b_{10}$) by at least two orders of magnitude -- much lower than any reasonable models or expectations in the ISM. Further, even if this somehow happened, it does not necessarily mean the disks would not be magnetically-dominated, but instead, with weaker fields (such that e.g.\ $|B_{\phi}| \ll |B_{\phi}^{\rm BP07}|$ in Eq.~\ref{eqn:BP07.compare}) could lead to conditions more like the MRI-dominated \citet{begelman.pringle:2007.acc.disks.strong.toroidal.fields} analytic disk models.

Formally speaking, there are of course physical criteria or conditions {\em implicit} in our assumption of how the magnetic fields are amplified (e.g.\ the models we consider with $B^{2} \propto \rho^{\gamma}$ where $1\lesssim \gamma \lesssim 2$ or $B \propto 1/H$). This includes, for example, the assumption that turbulent damping or buoyant escape of $B$ will not strongly damp the midplane magnetic fields (compared to their rapid growth/supply via advection and flux-freezing). These and related qualifications are tested and discussed explicitly in the simulations at length in \papertwo\ (see e.g.\ the discussion in \S~6.6 as well as \S~5.2 \&\ 9.5 therein), where it is also shown that the analytic models here are internally self-consistent in these respects. Given that we simply adopt these relations here, we refer interested readers to the more detailed tests therein.

So given these considerations and the weak mass dependence, we expect our assumptions and model conditions to reasonably apply to essentially the entire SMBH mass range $\sim 10^{4}-10^{10}\,{\rm M_{\odot}}$ at high accretion rates ($\dot{m} \sim 1$). The upper limit (in accretion rate) to the validity of the assumptions above may be set by the stability condition, $\dot{m} < 3000\,m_{7}^{-1/4}$ -- still allowing for extremely high accretion rates. The lower limit is less obvious but is likely set by similar physics as with a SS73-type disk; as $\dot{M}$ decreases, so does $\rho_{\rm mid} \propto \Sigma_{\rm gas}/H \propto \dot{m}$, and the disk becomes more optically-thin and lower-density. Recall that even at $\dot{m}\sim 1$, $\Sigma_{\rm gas}$ is significantly lower than a SS73 disk with the same $\dot{m}$, so this means the disks should approach the RIAF-type limit  at some $\dot{m} \ll 1$ where the cooling becomes too inefficient to maintain the temperatures and dissipate the gravitational energy of the inflowing gas. This will lead to a completely different thermal structure of the disk. Naively comparing the optically-thin H+He cooling rates for a fully-ionized plasma to the values implied in the scalings above, one finds that this still requires $\dot{m} \ll 0.1$, but more detailed models would be needed in this limit.

\subsection{What Drives the Turbulence?}
\label{sec:driving}

In this paper, as in e.g.\ SS73, we are intentionally agnostic to the source of the turbulence, but simply assume it is trans-\Alf{ic}, motivated in part by the simulations in \papertwo\ and by simple equipartition/non-linear saturation arguments. As discussed above, our conclusions are generally independent of this source, so long as it can sustain broadly order-of-magnitude trans-\Alf{ic} turbulence with \Alf\ Mach number $\mathcal{M}_{A}$ not {\em too} strongly dependent on radius $R$. In \papertwo, we study the turbulence in much more detail, demonstrating that these statements are true in the simulations, and that there are many viable drivers of the turbulence including (but not limited to): inward propagation of large-scale/global gravitational $m=1$ modes, gravito-turbulence in the outer disk, adiabatic compression/advection of turbulence with the accretion flow, and the \citet{pessah.psaltis:2005.mri.extensions.stronger.fields,das:2018.pessah.psaltis.limit.mri} ``Type II/III'' or ``SSMI/SHMI'' radial magnetic buoyancy instabilities. Other buoyancy instabilities may be present, though the Parker-type modes discussed in \citet{johansen.levin:2008.high.mdot.magnetized.disks} are likely suppressed owing to the combination of strong turbulence, low-$\beta$, and large $H/R$ making their wavelengths $\gtrsim R$ (\papertwo). Even the ``traditional'' linear MRI could operate in principle (despite the very low-$\beta$) in regions where $\partial \ln{B}_{\phi}/\partial\ln{R}$ is sufficiently close to $-1$.\footnote{More rigorously, where $|\partial\ln{B}_{\phi}/\partial\ln{R} + 1| < \beta\,(v_{\rm K}/v_{A})^{2}$ ($\lesssim 10^{-2}$, in the simulations in Fig.~\ref{fig:sim.model.comparison}).} However, it is difficult to uniquely separate and identify a single ``driver'' in such highly non-linear, multi-physics simulations. 

It is not even totally obvious that a turbulent driving mechanism is strictly necessary. For our purposes here, the angular momentum transport comes from some arbitrary combination of Maxwell+Reynolds stresses with $\nu_{\rm visc} \sim v_{A}^{2} /\Omega$. This can come from a trans-\Alf{ic} turbulent Reynolds stress, or turbulent Maxwell stress, but in \papertwo\ we show that in the inner disk (while there certainly {\em is} trans-\Alf{ic} turbulence present) this can actually be dominated by the mean-field $-\langle B_{\phi} B_{R}/4\pi \rangle$ Maxwell stress as the toroidal field amplifies from the radial field producing a $B_{\phi}-B_{R}$ anti-correlation giving rise to $\nu_{\rm visc} \sim v_{A}^{2}/\Omega$. We also discuss therein that the turbulence itself may be largely advected from larger scales, rather than strictly locally amplified. Of course, one could have additional angular momentum transport via effects like outflows, but the simulations in \papertwo\ find this is negligible on the scales we compare in Fig.~\ref{fig:sim.model.comparison}.

Given this and our discussion of different variations for $\psi_{A}$ and $\zeta$ above, it is worth discussing how different proposed saturation mechanisms for {\em either} the turbulence or toroidal-radial mean field might scale. Since instabilities like the SSMI/SHMI are powered by the mean field magnetic tension, non-linear saturation around equipartition ($\mathcal{M}_{A} \sim 1$) seems reasonable. But \citet{begelman:2023.mri.saturation.estimates} propose some alternative analytic hypotheses for the saturated amplitudes of the MRI and/or Type III/SHMI instabilities in strongly-magnetized, toroidal-field dominated disks based on quasi-linear theory, or we could analogously speculate that saturation occurs when the linear growth time of these instabilities becomes comparable to other timescales such as the weak turbulence cascade time or the gas accretion timescale. We have re-derived our model scalings assuming each of these saturation scenarios in turn, for any given $\gamma=4/3-5/3$, assuming either the SHMI or MRI dominates the driving: these are all equivalent to very small changes in $\zeta$ (compared to our ``default'' trans-\Alf{ic} assumption) within the range $-0.05 \lesssim \delta \zeta \lesssim 0.07$. Notably, this range of $\zeta$ is much smaller than the {\em ad-hoc} variations we already discussed above and showed do not change our conclusions. Thus, for the purposes of a global disk solution, most plausible saturation scenarios produce similar conclusions and, given the uncertainties, we do not feel it helpful to speculate further here. Future work using idealized numerical simulations will help to better constrain this physics.

\section{Comparison to Other Accretion Disk Models}
\label{sec:compare}

\subsection{Comparison to a Shakura \&\ Sunyaev $\alpha$ Disk}
\label{sec:analytic.ss73}

The above differs from the classic \citet{shakurasunyaev73} (henceforth SS73) $\alpha$-model in many details, but the most important and fundamental is that SS73 assume a gas-pressure dominated disk ($H\approx c_{s}/\Omega$) with $\alpha \approx (v_{t}/c_{s})^{2} \sim {\rm constant} < 1$. This in turn leads to many other differences: for example the disk structure and accretion depends on the opacity structure, the disks are strongly vertically stratified, $\beta > 1$ in the midplane, $t_{\rm cool} \gg t_{\rm dyn}$, and the surface densities $\Sigma_{\rm gas}$ and Toomre $Q$ parameter can be very different. 

To gain some insight, consider the ``equivalent'' $\alpha$ parameter $\alpha_{\rm therm,\,equiv} \equiv \varpi_{\nu}\,v_{t}^{2}/c_{s}^{2}$ from our model above (\S~\ref{sec:analytic.structure}) -- defined as the $\alpha$ parameter in the SS73 model that would give rise  to the same accretion rate as our fiducial model for a given $c_{s}$ or $\Sigma_{\rm gas}$. Note that the thermal assumptions in \S~\ref{sec:analytic.thermal} are similar (where the disk is optically thick) so the zeroth-order temperature is similar. This means $\alpha_{\rm therm,\,equiv} \sim 1/\beta$ as defined above -- so we see $\alpha_{\rm therm,\,equiv} \gg 1$ -- i.e.\ the accretion is much faster, for a given $\Sigma_{\rm gas}$, than the fastest-possible SS73 model (this is clear in $\langle v_{R}\rangle$ in Fig.~\ref{fig:sim.model.comparison}). This in turn means that for a given $\dot{M}$ in steady state, $\Sigma_{\rm gas}$ at a given physical radius is lower in our fiducial model by a factor of $\sim (\alpha_{\rm SS} / \alpha_{\rm therm,\,equiv}) \sim \alpha_{\rm SS}\, \beta$ relative to the $\Sigma_{\rm gas}$ from an SS73 model with the same $\dot{M}$ and some $\alpha_{\rm SS} < 1$. We see this in the density plotted in Fig.~\ref{fig:sim.model.comparison}. Meanwhile $H/R$, $v_{A}/v_{\rm K}$, and $v_{t}/v_{\rm K}$ are larger in our fiducial model relative to SS73 by factors of $\gtrsim \beta^{-1/2} \gg 1$. At a given $r$, this means $Q$ (which is dominated by just the thermal contribution in the SS73 model) is larger in our fiducial model relative to SS73 by a factor $\propto \sigma_{\rm eff}/\Sigma_{\rm gas}$, i.e.\ $Q/Q_{\rm SS} \sim \beta^{-3/2}\,\alpha_{\rm SS}^{-1}$, also shown directly in Fig.~\ref{fig:sim.model.comparison}. For a canonical $\alpha_{\rm SS} \sim 0.1$, this can amount to a factor of $\sim 10^8$ increase in $Q$!

As a result, if one assumes a SS73 $\alpha$ disk with $\alpha \sim 0.1-1$, then even for $\dot{m}$ below/around the Eddington limit one obtains the well-known result that the disk should fragment ($Q_{\rm SS} \lesssim 1$) outside of a few hundred gravitational radii. For the conditions in the simulations shown in Fig.~\ref{fig:sim.model.comparison} ($m_{7} \sim 1$ and super-Eddington $\dot{M} \sim 20\,{\rm M_{\odot}\,yr^{-1}}$) the prediction would be $Q_{\rm SS} \lesssim 1$ outside of a radius as small as $\sim 10^{-5}\,$pc. Instead, we see in the simulations and  model that, at all radii out to a few pc, $Q \gg 1$ and even $Q_{\rm thermal} \gtrsim 1$ (where we do not include magnetic or turbulent support but still see larger $Q$ because of the greatly reduced surface densities).

This also makes it clear that the simplified models we propose are not technically $\alpha$ models, since the effective $\alpha$ is not constant, though it does vary relatively weakly.

One non-intuitive result, comparing our default models to SS73 in Fig.~\ref{fig:sim.model.comparison}, is that the rms magnetic field strength {\em in absolute units} is actually {\em larger} in an SS73 disk (assuming $v_{t} \sim v_{A}$)! Indeed, computing the ratio from the analytic expressions above and from SS73:\footnote{We formally estimate the magnetic properties from SS73 by assuming $\alpha$ primarily arises from Maxwell stress, appropriate for e.g.\ MRI-driven transport. Adding e.g.\ a mean poloidal field to SS73 does not change our discussion.} $B_{\rm SS73}/B \sim 12\,m_{7}^{9/80}\tilde{r}^{1/48}\alpha_{0.1}^{1/20}\dot{m}^{-3/40}r_{{\rm ff},\,5}^{-1/16}$ -- i.e.\ an order of magnitude stronger fields appear in an SS73 disk, with extremely weak dependence on any parameter. But of course the models here feature much larger \Alf\ speed ($v_{A} \equiv |{\bf B}|/\sqrt{4\pi\rho}$) and orders-of-magnitude lower $\beta \sim  P_{\rm thermal}/P_{\rm B} \sim c^{2}_{s}/v^{2}_{A}$ compared to SS73: $v_{A} \gg c_{s}$ in the models here, while $v_{A} \ll c_{s}$ in SS73. But because of the enormous difference in density of the models (where $\rho$ and $P_{\rm thermal}$ in SS73 are larger than here by factors $\sim 10^{6}$), even a relatively large $|{\bf B}|$ in SS73 (needed to explain some $\alpha \sim 0.1$) still gives very low $v_{A} \ll c_{s}$ and therefore negligible magnetic disk support. This difference in $|{\bf B}|$, however, has important consequences for formation of these disks, helping to emphasize the point made in \papertwo\ that the  magnetic flux from the ISM needed to supply these disks is not particularly large (corresponding to completely ``mundane'' ISM field strengths). This is also important to keep in mind when considering potential observable signatures.

\subsection{Comparison to the Begelman \&\ Pringle Model}
\label{sec:analytic.bp07}

To our knowledge the closest analytic model in the literature (of those which attempt to actually predict the scalings of $B(R)$ in the disk) to that proposed herein is the model discussed in \citet{begelman.pringle:2007.acc.disks.strong.toroidal.fields} (hereafter BP07), which assumes an $\alpha$-like model for a disk with primarily toroidal magnetic fields and $\beta \ll 1$. While there are a number of differences in detail regarding some of the assumptions above (e.g.\ regarding the thermal structure, accounting for turbulence and its effects on stratification, etc.), these mostly have weak effects on the qualitative behaviors and conclusions of the models. 

The most important physical difference between the models proposed here and BP07 is the assumption of what sets the magnetic field strength. \papertwo\ shows that in the simulations motivating our assumptions, this appears to be given by simple flux-freezing and advection of ISM magnetic flux, which gives rise to an analytic scaling broadly similar to $|B_{\phi}| \propto \rho_{\rm mid}^{2/3}$ or $\propto H^{-1}$, and introduces the concept of the free-fall radius above because the magnetic flux, and hence the disk properties, explicitly depend on the outer boundary conditions (the magnetic flux being fed into the disk). In BP07, the authors essentially assume the opposite limit, where the ``seed'' or ``initial'' field is some small/trace field, which is then amplified by the MRI, until the MRI saturates and its linear growth ceases at approximately the threshold estimated in \citet{terquem:1996.toroidal.field.bouyancy.instability,pessah.psaltis:2005.mri.extensions.stronger.fields}:\footnote{More generally allowing for radial stratification, one should take this limit to be $v_{\rm sat,\,mri} \sim \sqrt{c_{s}\,v_{\rm K}/|1 + \partial \ln{B}_{\phi}/\partial \ln{R}|}$ following \citet{pessah.psaltis:2005.mri.extensions.stronger.fields}, but this amounts to a relatively small systematic factor of $1.7$ in  Eq.~\ref{eqn:BP07.compare}, and in BP07 the simpler scaling $v_{A} \sim v_{\rm sat,\,mri} \sim \sqrt{c_{s}\,v_{\rm K}}$ was used so we compare it here.} $v_{A} \sim v_{\rm sat,\,mri} \sim \sqrt{c_{s}\,v_{\rm K}}$.

As shown explicitly in \papertwo\ in multiple examples, $v_{A} \sim v_{\rm sat,\,mri}$ is not, in fact, a good description of the magnetic fields in the simulations: their amplitude is everywhere order-of-magnitude {\em larger} than this threshold $v_{\rm sat,\,mri}$ ($v_{A} \gg v_{\rm sat,\,mri}$ or equivalently $|{\bf B}|_{\rm simulation} \gg B_{\rm sat,\,mri}$), they obey a different power-law scaling with BH-centric radius and density, and do not exhibit any clear dependence on sound speed at a given radius and density (unlike $v_{\rm sat,\,mri}$ which depends directly on $c_{s}$). But as shown in \papertwo, all of these results are a natural expectation if the field strength determined by flux-freezing and the advected field is already above this nominal critical value (though it is also possible that MRI growth continues in more complicated, non-linear, radially-stratified systems like those here). We can validate this assumption in our analytic model above, calculating the ratio at each radius of the predicted field strength $|B_{\phi}|$ to this threshold value $B_{\rm BP07} \equiv B_{\rm sat,\,mri}$ (which should be $\gtrsim 1$):
\begin{align}
\label{eqn:BP07.compare} \frac{|B_{\phi}|}{|B_{\phi}^{\rm BP07}|} &\sim \frac{v_{A}}{\sqrt{c_{s}\,v_{\rm K}}} 
\sim 20\,\frac{m_{7}^{1/8}\,f_{\tau}^{1/16}\,\tilde{r}^{5/48}}{\dot{m}^{1/16}\,r_{{\rm ff},5}^{1/16}} \\
\nonumber &\sim 4\,\frac{f_{\tau}^{1/16}\,m_{7}^{11/48}\,x_{g}^{5/48}}{\dot{m}^{1/16}\,r_{{\rm ff},5}^{1/6}} 
\nonumber \sim 20\,m_{7}^{0.09}\,\dot{m}^{-0.06}\,\tilde{r}^{0.1}.
\end{align}
So indeed we predict order-of-magnitude larger fields, and this conclusion is relatively insensitive to the conditions and other detailed assumptions above.\footnote{The more general version allowing for different $\psi_{A}$ and $\zeta$ is $v_{A}/\sqrt{c_{s}\,v_{\rm K}} \sim 20\,\psi_{A}\,\tilde{r}^{1/112 + 2\zeta/7}$. Scaling to the stellar-mass regime with the units above gives $| B_{\phi} |/|B_{\phi}^{\rm BP07}| \sim 2\,(M_{\rm bh}/10\,{\rm M_{\odot}})^{25/144}\,(r/6\,R_{g})^{5/48}\,\dot{m}^{-1/16}\,(P_{\rm ff}/{\rm day})^{-1/9}$. Using the different flux-freezing closure $B\propto H^{-1}$  \citep{machida:2006.mag.supported.disks.from.cooling.around.adafs} gives nearly-identical scalings.}

Despite this, most of our qualitative conclusions are similar to BP07 -- indeed, any magnetically-dominated model where $v_{t}/v_{A}$ and $v_{A}/v_{\rm K}$ do not vary {\em too} strongly as a function of $r$ would give many broadly similar conclusions to our proposed model regarding the scalings of $Q$, $\beta$, $\Sigma_{\rm gas}$, etc. And while the detailed power-law dependencies are different, the ``weak'' dependencies remain ``weak'' (e.g.\ terms scaling here to the $1/16$ power might scale as $1/36$ or $1/12$ in BP07). 

Nonetheless, there is one important consequence of the order-of-magnitude difference as $\tilde{r} \rightarrow 1$ in Eq.~\eqref{eqn:BP07.compare}. Recall, the effective ``$\alpha$'' scales as $\propto v_{t}^{2} \propto \beta^{-1} \propto B^{2}$, and the effective $\Sigma_{\rm gas}$ scales as $1/\alpha$, and $Q \propto \sigma_{\rm eff}/\Sigma_{\rm gas} \propto \alpha^{3/2}$. So at small radii (approaching the ISCO), the difference with BP07 is not so dramatic, but at large radii -- noting the ratio in Eq.~\eqref{eqn:BP07.compare} increases with $r$ -- a factor of $\sim 10$ in magnetic flux translates to factor $\sim 100$ lower $\Sigma_{\rm gas}$ and $\sim 1000$ larger $Q$. Basically, in BP07, $\Sigma_{\rm gas}$ decreases more slowly (relative to our model here) at large $r$ because $v_{t} \propto \sqrt{c_{s}\,v_{\rm K}}$ declines more rapidly. As BP07 note, this leads to a ``bottleneck'' where they predict their disk models would fragment before reaching the BHROI (roughly equal to our $r_{\rm ff}$ here) for a critical $\dot{m} \lesssim 1.8\,T_{70}^{3/4}\,m_{7}^{-5/8}$ (where $T_{70}$ scales the temperature to $\sim 70$\,K allowing even for the temperature at large radii to be set by complex ISM physics making it warmer than predicted by the blackbody scaling, as discussed above and in BP07, and we rescale from their definition of $\dot{m}$ to ours). Whereas we see from Eq.~\eqref{eqn:Q} (and directly in the simulations) that $Q > 1$, and even $Q_{\rm thermal} > 1$, out to the BHROI for $\dot{m}$ as large as $\gtrsim 1000$.

In short, while the BP07 model is already much less prone to fragmentation than a thermal-dominated model like SS73, by allowing for  larger fields set by  flux-freezing  at larger radii (as opposed to only the field strengths at which the MRI saturates according to local linear theory in a non-stratified disk), our model allows the disk to support factor $\sim 1000$ larger inflow rates at all radii from the BHROI inwards. 

\subsection{Comparison to Magnetically Elevated or Arrested Disks}
\label{sec:analytic.arrested}

In \papertwo, we discuss and demonstrate extensively how the simulation disks are qualitatively and quantitatively different in their basic properties from historical models of magnetically ``levitated'' or ``elevated'' or ``arrested'' disks (see e.g.\ \S~9 therein). Rather than repeat all of the discussion and detailed comparisons therein, we briefly summarize how this also applies to the analytic models here. 

First, consider levitated or elevated disks. The defining feature of these models \citep[see e.g.][]{miller.stone:2000.magnetically.elevated.disk} is that the magnetic fields are weak ($\beta \gg 1$) in the disk midplane (where the structure is akin to an SS73-type disk), but rise in importance at larger vertical heights above the disk until $\beta \lesssim 1$ at a few scale-heights ($|z| \gg H$). Clearly this is opposite our fundamental assumptions for $\beta$ and the disk midplane structure, and produces predictions for disk properties like $\rho$, $v_{r}$, $v_{t}/v_{\rm K}$, $Q$, etc.\ akin to SS73 and completely different from our model in Fig.~\ref{fig:sim.model.comparison}. Moreover in \papertwo\ we show that the simulation disks are actually both weakly and inversely stratified ($\beta$ is lowest, not highest, in the midplane) relative to the elevated assumption.

Second, consider magnetically arrested (MAD) disks, defined by dominant mean large-scale vertical/poloidal magnetic fields (${\bf B} \approx \langle B_{z} \rangle\,\hat{z}$) whose magnetic tension balances gravity and prevents accretion \citep{bisnovatyi.kogan:1976.mad.disk,narayan:2003.mad.disk}. It is easy to verify that our disk models never satisfy the standard MAD criterion  of $\langle B_{z} \rangle_{\rm mad}^{2}/8\pi \gtrsim G\,M_{\rm bh}\,\Sigma_{\rm gas}/4\,R^{2}$, even if we assumed the field was entirely in a coherent mean poloidal configuration. To see this, note the criterion can be re-arranged (using our implicit assumption for magnetically-dominated disks that $H/R \approx v_{A}/v_{\rm K}$) as $\langle B_{z} \rangle_{\rm mad}^{2} / |{\bf B}|^{2} \gtrsim v_{\rm K}/v_{A}$ -- which can only be realistically satisfied if both $v_{A} \gtrsim v_{\rm K}$ (the \Alf\ speed rises to be super-Keplerian) and $|\langle B_{z} \rangle| \approx |{\bf B}|$ (the field is primarily in a coherent mean vertical component). As noted above, even ignoring the field geometry considerations, achieving the former requires much more extreme amplification than seen in the simulations/assumed here ($\gamma \gtrsim 2$, pushing towards the upper limit of what is possible for perfect flux-freezing in a laminar cylindrical vertical-field-only  disk), together with the assumption that the turbulence becomes weak ($v_{t} \ll v_{A}$) rather than remaining trans-\Alf{ic} (since in the trans-\Alf{ic} limit with such strong amplification, the solutions would extrapolate to free-fall, promoted by Maxwell+Reynolds stress, rather than arrested/static solutions supported by magnetic tension). Moreover, as  discussed in \papertwo\ at length, in the simulations motivating our models here, the fields are primarily toroidal, with secondary radial and quasi-isotropic turbulent components (so the mean vertical field $|\langle B_{z} \rangle| \ll |{\bf B}|$ is generally the weakest component of ${\bf B}$). This has important qualitative effects, changing the field amplification, role of magnetic pressure setting the vertical height and midplane structure of the disk, and the sign of the magnetic torques/Maxwell stresses (making the fields promote accretion, rather than suppress it) -- all of which are implicit in our model assumptions. And unless one begins from boundary conditions with ${\bf B} \approx \langle B_{z} \rangle\,\hat{z}$, it is essentially impossible to generate this state via amplification within the disk (one cannot generically amplify the mean vertical field from trace to stronger than the turbulent field $|\langle B_{z} \rangle| \gg \langle | \delta B_{z} | \rangle$). 

Finally, as discussed in detail in \papertwo, although the criterion for a disk becoming MAD-like ``eventually'' is sometimes stated in terms of a critical magnetic flux at some arbitrarily large radius (e.g.\ some $\Phi_{\rm crit} \sim \pi\,R^{2}\,|{\bf B}|_{\rm crit} \sim {\rm \mu G\,pc^{2}}$, as opposed to the local MAD criterion discussed above), this scaling makes a number of further assumptions -- in {\em addition} to both the strong amplification ($\gamma \gtrsim 2$) and mean-vertical field dominated (${\bf B} \approx \langle B_{z} \rangle\,\hat{z}$) assumptions above -- which are not valid in our model. Most notably, it also assumes that the field plays a negligible role in the disk structure (outside of the ``arrested'' radius) and so the disk can be treated as a $\beta \ll 1$ SS73-like $\alpha$-disk, which we have shown is dramatically different from our model predictions. So while it is certainly possible that some accretion disks could, in general, approach a MAD-like state, it requires {\em qualitatively} different boundary/initial conditions, field amplification, and accretion disk structure from what we assume and model herein.

\subsection{Comparison to Pariev et al.\ 2003}
\label{sec:analytic.p03}

We note that the power-law scalings with radius (logarithmic slopes) that we obtain represent a generalization of those derived in \citet{pariev:2003.mag.dominated.disk.models} (P03): we obtain their scalings if we restrict to trans-\Alf{ic} turbulence, neglect advective radiation transport (assume $f_{\tau} = \tau^{-1}$), and set $\gamma \rightarrow \delta_{\rm P03}/(3 \delta_{\rm P03}-3)$ and $\zeta \rightarrow 3-2 \delta_{\rm P03}$ in terms of their arbitrary parameter $\delta_{\rm P03}$ defined by their assumption that $B_{\rm turb}(R) = B_{10,\,{\rm P03}} (R/20 R_{g})^{-\delta_{\rm P03}}$. This is despite the fact that P03 considered very different physical conditions (e.g.\ much lower accretion rates) and motivations, and assumed only ``pure turbulent'' magnetic fields (negligible flux freezing and mean fields). This similarity arises because the trans-\Alf{ic} assumption gives, by definition, $|\delta {\bf B}^{2}|^{1/2} \propto |\langle {\bf B} \rangle |$, so should produce the same dimensional scalings for appropriate choice of $\zeta$ and $\gamma$. But in addition to extending what is presented in P03 by varying these assumptions and connecting the disks to their outer boundary at $r_{\rm ff}$ or the BHROI, our model attempts to actually predict quantities like $\delta_{\rm P03}$ and $B_{10,\,{\rm P03}}$ (which are left arbitrary in P03) from the simulations and physical arguments. It is also worth noting that P03's discussion, which considered the absolute normalization of the magnetic field (in Gauss) around the ISCO required by the self-consistency of their assumptions, does not explain why $B_{10,\,{\rm P03}}$ would or should naturally lie in their ``valid'' range for a given $\dot{M}$ and $M_{\rm BH}$. By introducing the outer boundary conditions for the disk, we show that this is automatically ensured so long as $\psi_{A}$ is not too small (\S~\ref{sec:analytic.caveats}).

\section{Limiting Regimes of the Models and Behavior at Small Radii}
\label{sec:analytic.eddington}

The most basic requirement for accretion disks to be in the state predicted here is that their outer boundary conditions satisfy our assumptions: namely, gas is accreting which is relatively thermally ``cold'' ($c_{s} \ll v_{\rm K}$) and well-magnetized ($\beta \lesssim 1$) with (as detailed in \papertwo) a steady supply of toroidal and radial magnetic flux. Beyond this, as noted above, at sufficiently large $\dot{m}$, the outer disks predicted here can still become Toomre unstable, but this will not occur until $\dot{m} \gtrsim 3000\,m_{7}^{-1/4}$. At sufficiently low $\dot{m}$ (probably $\dot{m} \ll 0.1$), the thermal structure may be modified by inefficient cooling. Provided these conditions are met, the models here are designed to be extrapolated out to the BHROI, beyond which point, by definition, the physics must change since the potential changes, star formation becomes non-negligible, the thermal structure is set by diffuse ISM heating and cooling processes rather than accretion heating, etc. And indeed the simulations to which we compare are focused on this outer disk, from the BHROI at $x_{g} \sim 10^{7}$ (and beyond) to $x_{g} \sim 600$. But the behavior in the inner disk down to the ISCO/horizon ($x_{g} \sim$\,a few) is less obvious.

Approaching the horizon, one has the boundary condition $v_{r} \sim v_{\rm K} \sim c$, while our ``default'' solution (Eq.~\ref{eqn:veff.model}) predicts $v_{r}/v_{\rm K} \rightarrow (R_{g}/R_{\rm BHROI})^{1/3} \sim (\sigma_{\rm gal}/c)^{2/3} \ll 1$. Thus there would have to be some transition in physics between the radii of interest here and the near-horizon or ``free-fall'' region (i.e.\ one limitation of the analytic models here is that they must be modified near the inner boundary/horizon), but that is not particularly surprising (and similarly holds if one attempts to extrapolate most $\alpha$-disk models, including SS73 or BP07, from large radii to the horizon). Interestingly, the simpler solution for slightly ``stiffer'' magnetic amplification ($\gamma=5/3$; Eqs.~\eqref{eqn:veff.stiff}-\eqref{eqn:Q.stiff}) extrapolates  to $v_{r} \sim v_{\rm K}$ (for natural order-unity $\psi_{A}$). Thus, the transition in the inner disk could simply amount to slightly more efficient field amplification. 

At high accretion rates approaching or exceeding Eddington ($\dot{m} \gtrsim 1$), it is also important to examine the role of radiation pressure in the disk. For the models in \S~\ref{sec:analytic}, the ratio of midplane radiation pressure $P_{\rm rad} \sim (1/3)\,a\,T^{4}$ to ``other'' (thermal+magnetic+turbulent) pressure $P_{\rm other}$ is just $P_{\rm rad}/P_{\rm other} \sim f_{\tau}^{-1}\,(2\,v_{t}^{2}/(v_{t}^{2}+v_{A}^{2}))\,v_{A}/c \sim f_{\tau}^{-1}\,v_{A}/c \sim f_{\tau}^{-1}\,v_{t}/c$ (see \S~\ref{sec:analytic.thermal}). If we assume the standard optically thick diffusive-radiation-transport limit, $f_{\tau}^{-1} \sim \tau \sim \Sigma_{\rm gas}\,\kappa$, then this is negligible in the outer disk: $P_{\rm rad}/P_{\rm other} \sim 10^{-6} \,\dot{m}\, (m_{7}/r_{{\rm ff},\,5})\,(\kappa/\kappa_{\rm es})\,\tilde{r}^{-7/6}$ (in addition, for the conditions in Fig.~\ref{fig:sim.model.comparison}, the outer disk is primarily atomic, so $\kappa \ll \kappa_{\rm es}$). But $P_{\rm rad}/P_{\rm other}$ will increase inwards until it crosses unity at some\footnote{If we use the alternative ($\gamma=5/3$) scalings from  Eqs.~\ref{eqn:veff.stiff}-\ref{eqn:Q.stiff} to caculate where $P_{\rm rad}/P_{\rm other}$ would extrapolate to $>1$ assuming diffusive transport ($f_{\tau}^{-1}=\tau$), we obtain the similar result $x_{g} \lesssim 100\,(0.3/\psi_{A})\,(\kappa/\kappa_{\rm es})\,\dot{m}$, noting the best-fit $\psi_{A} \sim 0.3$ from this model to the scalings in Fig.~\ref{fig:sim.model.comparison}.} $x_{g} \lesssim 150\,(r_{\rm ff,\,5}/m_{7})^{1/7}\,(\kappa/\kappa_{\rm es})^{6/7}\,\dot{m}^{6/7}$. Note this is much is closer to the horizon than what one would obtain for a standard SS73 disk (where $P_{\rm rad} \gtrsim P_{\rm other}$ at $x_{g}^{\rm SS73} \lesssim 9000\,\dot{m}^{0.8}\,(\alpha m_{7})^{0.1}\,(\kappa/\kappa_{\rm es})^{0.9}$) owing to the much stronger magnetic pressure. But it is still well outside the horizon for large $\dot{m}$. Interior to this radius, naively the solutions would be modified (with radiation pressure ``puffing up'' the disk and setting $H$), and we could derive an appropriately modified solution which retains the key characteristics of our models herein (akin to the ``region (a)'' model in SS73). However, at even smaller radii, approximately (up to an $\mathcal{O}(1)$ constant) where the disk luminosity $L_{\rm disk} \sim \pi\,R^{2}\,\sigma_{B}\,T_{\rm eff}^{4}$ exceeds $L_{\rm Edd}$, it becomes impossible to support the disk even with $H \sim R$ and the radiation pressure forces exceed gravity, {\em if} the heat transport is diffusive with electron-scattering opacity (the minimum scale height effectively scales as $(H/R)_{\rm min} \sim (f_{\tau}^{-1}/\tau_{\rm es})\,(L_{\rm disk}/L_{\rm Edd})$, with $\tau_{\rm es} \equiv \Sigma_{\rm gas}\,\kappa_{\rm es}$). For $f_{\tau}^{-1} \sim \tau_{\rm es}$, this would occur at $x_{g} \sim 4\dot{m}$, well outside the horizon/ISCO for $\dot{m} \gg 1$. 

However, the ratio $P_{\rm rad}/P_{\rm other}$  in the diffusive limit ($f_{\tau}^{-1} \sim \tau$) in the models above is just $\sim v_{t}/(c/\tau)$ -- i.e.\ the same as the ratio of the advective ($\sim v_{t}$) to diffusive ($c/\tau$) heat transport rates. So in the models here, at the same radii where $P_{\rm rad}$ would naively begin to exceed $P_{\rm other}$, the (vertical/local) heat transport becomes advective ($v_{\rm adv} \sim v_{t} \gtrsim v_{\rm diff} \sim c/\tau$), so the disk should become vertically mixed and weakly stratified with $f_{\tau}^{-1} \sim 1 \ll \tau$. If we assume the maximum-possible stratification scales as $f_{\tau}^{-1} \sim c/v_{t}$, the disk would converge to $P_{\rm rad} \sim P_{\rm mag} \sim P_{\rm turb} \gg P_{\rm thermal}$ (if instead $f_{\tau}^{-1} \rightarrow 1$, then the disk remains with $P_{\rm mag} \sim P_{\rm turb} > P_{\rm rad} \gg P_{\rm thermal}$ at all $r$) and the solutions we describe above would only be weakly modified by a systematic $\mathcal{O}(1)$ prefactor, and so $H < R$ (from the original solutions) and continued accretion could be sustained down to near-horizon scales. Note that {\em if} we continue to assume diffusive radiation transport at $v_{\rm diff} \sim c/\tau$, then the radial advective speed $v_{R} \sim v_{t}^{2}/v_{\rm K}$ becomes larger than the diffusive speed at $x_{g} \lesssim (10-20)\,(\kappa/\kappa_{\rm es})\,\dot{m}$, suggesting the system would become radiatively inefficient akin to the standard slim disk solution. However, because we just argued the radiation is not ``locked'' to the diffusive speed at these radii and $v_{\rm t} > v_{R}$ always, the actual radiative efficiency of the inner disk when $\dot{m} \gg 1$ (and thus whether the quasars can appear super-Eddington or not, in the observable luminosity sense) will be set by some non-linear competition between accretion (radial advection) and convective/turbulent escape (vertical advection) of radiation.

Given the somewhat speculative discussion above, it clearly would be valuable to directly simulate disks like those proposed here on scales smaller than those resolved in the \papertwo\ simulations, resolving the radii where the key transitions above should take place with explicit radiation-magnetohydrodynamics, in order to properly model and understand whether the heat transport becomes advective (in both the vertical and/or radial directions) and how this modifies the disk structure, thermal properties, and radiative efficiencies. This could be done in the Newtonian or pseudo-Newtonian limits for sufficiently large $\dot{m}$, given the characteristic $x_{g}$ above. But obviously, properly modeling scales of order the horizon (the inner boundary) would require GRMHD simulations.

\section{Summary \&\ Conclusions}
\label{sec:conclusions}

We present a simple analytic similarity model for strongly-magnetized flux-frozen accretion disks; specifically, disks with midplane $\beta \ll 1$, super-sonic turbulence, and efficient cooling. We validate this model against recent numerical simulations that found such disks forming around SMBHs accreting at quasar-level accretion rates from cosmological initial conditions, and show that it can reasonably explain the qualitative behaviors and scalings in the simulations.

We show that in this limit, for reasonable assumptions (e.g.\ trans-\Alf{ic} turbulence), the model is internally consistent and entirely specified by the boundary conditions, specifically the accretion rate and magnetic flux around the ``free fall'' radius (the outer radius at which gas is first captured into the accretion disk, and the model maps to a solution of free-fall onto the SMBH, which should occur near the BHROI). In the outer accretion disk, which extends all the way to the BHROI (exterior to which the physics must become more ISM-like), we show these disks should remain consistent and maintain gravitational stability ($Q \gg 1$) up to extremely large Eddington-scaled accretion rates $\dot{m} < 3000\,(M_{\rm bh}/10^{7}\,{\rm M_{\odot}})^{-1/4}$. We show that most of the predicted structure and scalings are insensitive to the detailed scalings adopted for the magnetic field structure or other parameter choices, provided the limits above are satisfied (and a sufficient supply of coherent toroidal/radial magnetic flux exists in the first place).

These disks have qualitatively different structure from a \citet{shakurasunyaev73}-like $\alpha$ disk, most notably with $Q$ larger by factors up to $\sim 10^{8}$ for the simulations to which we compare. They are also qualitatively distinct from magnetically ``elevated'' or ``arrested'' disks. They share similarities with the model outlined in \citet{begelman.pringle:2007.acc.disks.strong.toroidal.fields}, but differ in some key respects, most notably in that the magnetic fields set by accretion of ISM magnetic flux (from typical ISM field strengths) can be much larger than the upper limit assumed in \citet{begelman.pringle:2007.acc.disks.strong.toroidal.fields} (which takes the field to arise from the local unstratified, weak seed-field MRI only). This difference  leads to different scalings  and allows factors of $\gtrsim 1000$ larger accretion rates to be stably supported. Indeed, as discussed in \papertwo\ extensively, a remarkable feature of the simulations (and implicit in these analytic models) is that such hyper-magnetized disks do not strictly need to be ``powered'' by the MRI, and can feature magnetic fields much stronger than the often-quoted upper limit for linear MRI growth from  \citet{pessah.psaltis:2005.mri.extensions.stronger.fields}. 

Finally, we discuss implications for the inner disk and regimes where this model may require extensions at either much lower accretion rates or much smaller radii (approaching the ISCO). At small radii, our model, which has been primarily focused on the outer disk given the simulations and questions of gravitational stability,  may require modification due to general relativistic effects and strong radiation pressure if $\dot{m} \gg 1$ (where we argue the heat transport may become advective, enabling sustained super-Eddington accretion). This will be an important subject for future work. It will also be important to explore the limits of the boundary conditions (in terms of accretion rate and magnetic flux) where such disks can be sustained.

That said, the robust nature of the analytic scalings we derive and their relative insensitivity to the details of the boundary conditions, BH mass, and accretion rates, together with the fact that the first (and thus far only) cosmological radiation-MHD simulations to resolve quasar accretion disks saw such hyper-magnetized disks form naturally, all suggests such disks may be common around quasars in nature. Given that they naturally appear to resolve the decades-old problem of gravitational instability in the outer accretion disk, it seems crucial to study their properties in more detail. Their spectral properties, detailed internal structure, and behaviors at small radii and at extreme (low and high $\dot{m}$) regimes should all be explored more thoroughly in numerical accretion disk simulations. The key difference between such disks and the overwhelming majority of historical work on accretion disk simulations is the boundary/initial condition assumed for the magnetic flux (discussed in detail in \papertwo) -- we discuss how this is qualitatively distinct from e.g.\ models of magnetically arrested or elevated disks. Our hope is that the simple analytic models in this paper can inform and guide such studies, providing some guidance for extrapolation, but also be refined and improved by such studies in the future.

\begin{acknowledgements}
Support for PFH was provided by NSF Research Grants 1911233, 20009234, 2108318, NSF CAREER grant 1455342, NASA grants 80NSSC18K0562, HST-AR-15800. JS acknowledges the support of the Royal Society Te Ap\=arangi, through Marsden-Fund grant  MFP-UOO2221 and Rutherford Discovery Fellowship  RDF-U001804. CAFG was supported by NSF grants AST-2108230, 1652522; NASA grants 17-ATP17-0067, 21-ATP21-0036, HST-GO-16730.016-A, TM2-23005X.
\end{acknowledgements}

\bibliographystyle{mn2e}
\bibliography{ms_extracted}

\end{document}